\documentclass[onecolumn,3p,times,procedia]{elsarticle}

\usepackage{ecrc}
\usepackage{graphicx}
\usepackage{titlesec}
\usepackage{subfigure}
\usepackage{multirow}
\usepackage{diagbox}
\usepackage{threeparttable}
\usepackage{color}

\volume{00}

\firstpage{1}

\runauth{}

\jid{procs}

\CopyrightLine{2022}{Published by Elsevier Ltd.}

\usepackage{amssymb}
\usepackage{amsmath}
\usepackage{multicol}

\usepackage[figuresright]{rotating}
\usepackage{bm}
\usepackage{caption}

\usepackage{fancyhdr}
\fancyheadoffset[RO,EL]{0pt}
\usepackage{geometry}

\begin{document}

\begin{frontmatter}




\dochead{}
\title{
\begin{flushleft}
{\bf \Huge Deep learning for joint channel estimation and feedback in massive MIMO systems}
\end{flushleft}
}
 %

\author[]{\bf \Large \leftline {Jiajia Guo$^a$, Tong Chen$^a$, Shi Jin$^a$$^*$, Geoffrey~Ye~Li$^b$, Xin Wang$^c$,}
\leftline {Xiaolin Hou$^c$}}

\address{\bf  \leftline {$^a$National Mobile Communications Research Laboratory, Southeast University, Nanjing 210096, P. R. China.}

\bf  \leftline {$^b$ Department of Electrical and Electronic Engineering, Imperial College London, London SW7 2AZ, U.K.}

\bf  \leftline {$^c$ DOCOMO Beijing Communications Laboratories Co., Ltd., Beijing, P. R. China}

}

\cortext[]{This is an earlier version.}

\begin{abstract}

The great potentials of massive Multiple-Input Multiple-Output (MIMO) in Frequency Division Duplex (FDD) mode can be fully exploited when the downlink Channel State Information (CSI) is available at base stations.
However, the accurate CSI is difficult to obtain due to the large amount of feedback overhead caused by massive antennas. In this paper, we propose a deep learning based joint channel estimation and feedback framework, which comprehensively realizes the estimation, compression, and reconstruction of downlink channels in FDD massive MIMO systems. 
Two networks are constructed to perform estimation and feedback explicitly and implicitly. 
The explicit network adopts a multi-Signal-to-Noise-Ratios (SNRs) technique to obtain a single trained channel estimation subnet that works well with different SNRs and employs a deep residual network to reconstruct the channels, while the implicit network directly compresses pilots and sends them back to reduce network parameters. 
Quantization module is also designed to generate data-bearing bitstreams. 
Simulation results show that the two proposed networks exhibit excellent performance of reconstruction and are robust to different environments and quantization errors.

\end{abstract}


\end{frontmatter}



\section{Introduction}
\label{section: introduction}

The massive Multiple-Input Multiple-Output (MIMO) technology has been regarded as one of the key technologies for Fifth-Generation (5G) and future cellular communication systems \cite{marzetta2015massive,wong2017key,dang2020should}.
Massive MIMO systems exploit a much larger number of degrees of spatial freedom than conventional MIMO systems through deploying massive antennas at the Base Station (BS) in a distributed or centralized way. 
Without increasing the spectrum resources and transmitting power, such a massive MIMO system can increase channel capacity by multifold and significantly reduce the interference among users by making full use of the spatial resources. 

However, the exploitation of the potential benefits provided by the massive MIMO technology for signal detection, beamforming, resource allocation, etc, is significantly affected by whether the accurate uplink and downlink Channel State Information (CSI) are known or perfectly acquired by the transmitter. 
In practical systems, channels are unknown to the transmitter and have to be estimated through pilots at the receiver first. 
In the Time Division Duplex (TDD) mode, channel reciprocity between the uplink and downlink is leveraged to directly attain the downlink CSI from the uplink CSI estimates. 
However, channel reciprocity does not always exist in the TDD mode and the corresponding calibration process is difficult and complicated, so that the acquired uplink CSI estimates may be inaccurate for the downlink \cite{6798744,rusek2012scaling}. 

In current cellular systems, the Frequency Division Duplexing (FDD) mode, where the uplink and downlink are operated at different carrier frequencies, is widely used. 
To obtain downlink CSI at the base station in the FDD systems where instantaneous channel reciprocity does not hold in general, the UE has to first estimate the downlink CSI via transmitted pilots and then send the estimated CSI back to the BS through feedback uplinks. 
The dimension of the feedback channel matrices is huge due to the large-scale transmit antennas, thus causing a prohibitively large feedback overhead. 
Although feedback overhead can be reduced by traditional methods, such as vector quantization or codebook-based approaches \cite{9426450,4641946}, it still increases proportionally with the number of transmit antennas. 
An alternative scheme should be found to reduce the feedback overhead in massive MIMO systems.

With the increase in transmit antennas at the BS, the channels will present the sparse property in certain domains due to limited local clusters around the BS \cite{qin2018sparse}. 
Therefore, Compressed Sensing (CS) based CSI feedback has been regarded as an advanced and promising method for compression and reconstruction of the CSI in massive MIMO systems \cite{8284057}. 
The large-scale channel matrices are randomly projected onto a lower-dimensional subspace by CS to get the compressed CSI, while the reconstruction is realized based on the compressed CSI. 
Many CS-based algorithms \cite{6214417,6816089,sim2016compressed,8802261} have been proposed to achieve the optimal reconstruction performance.
However, CS techniques rely heavily on the prior sparsity assumption of channels, adopt random projection matrix without making use of the channel statistics, and require a large amount of computational time and resources, therefore it cannot satisfy the requirement of real-time processing in practical deployment.
Hence, a feedback mechanism that can quickly and accurately recover CSI from low-dimensional measurements is urgently needed in massive MIMO systems.
The challenges of CSI feedback provide a strong impetus to the introduction of advanced technologies, such as deep learning (DL), into communication systems.

DL-based schemes have already been utilized to overcome the drawbacks and improve the performance of various traditional algorithms \cite{8233654, 8663966,liu2021toward} such as signal detection \cite{ye2018power}, channel estimation \cite{8353153,9410430}, channel prediction \cite{9277535}, and end-to-end transceiver design \cite{8985539}. 
With regard to CSI feedback, inspired by the great breakthrough made by Convolutional Neural Network (CNN) in CS image reconstruction \cite{7780424}, 
researchers in \cite{8322184} have proposed a DL based massive MIMO CSI compression and reconstruction scheme, called CsiNet, which adopts an unsupervised autoencoder architecture to mimic the CS process and outperforms CS based algorithms by a large margin \cite{mashhadi2020deep}. 
Then, the novel scheme \cite{8482358}, called CsiNet-LSTM, extracts both spatial and temporal correlation features to improve the reconstruction performance. 
The scheme in \cite{li2019spatio-temporal} utilizes a deep recurrent neural network to exploit temporal correlation, with depthwise separable convolution adopted to compress the model. 
The denoise network in \cite{9076084} deals with interference and nonlinear effect in the channel feedback process to improve performance.
In addition, in \cite{8972904}, quantization module is enrolled and an offset network based on residual learning is deployed at the receiver to counteract the quantization errors. 
Through modifying the structure and reusing Fully-Connected (FC) layers, the multiple-rate compression network in \cite{8972904} adapts to different environments without increasing parameters. 
Besides, the adaptive network in \cite{lu2019multi-resolution} extracts channel features on multiple resolutions to make the network robust to various environments.
A forged complex-valued input layer is proposed to process complex channel signals in \cite{9497358}.
Network pruning and quantization techniques have been investigated in \cite{Guo2019CompressionAA,9373670} to reduce the number of the network parameters in DL based CSI feedback scheme.
The DL-based joint design of feedback and beamforming/precoding is proposed in \cite{9279228}.

All the aforementioned DL based CSI feedback schemes are based on the assumption that the downlink channels are accurately known to the user, which is actually impossible in practical communication systems \cite{mashhadi2020deep}.
Downlink channel estimation must be carried out to attain the downlink CSI before sending the estimated CSI back and the performance of channel feedback relies heavily on that of channel estimation.  
However, the channel characteristics of massive MIMO systems are so complex that traditional channel estimation algorithms generally lack the capability of detecting the real-time variation of channel conditions and may consume large resources to accomplish nonlinear reconstruction procedures.

As a potential alternative, the DL technique can also be applied in channel estimation. 
In \cite{8353153}, a denoising neural network, unfolded from the conventional iterative approximate message passing algorithm, is utilized to estimate the transmitter-receiver channels of mmWave massive MIMO in the spatial domain (antenna space).
A deep CNN-based channel estimation framework has been proposed in
\cite{8752012} to exploit the temporal correlation of time-varying channels.
Moreover, a DNN architecture to jointly design the pilot signals and channel estimation module end-to-end has been proposed to avoid performance loss caused by separate design in \cite{ma2020data-driven}.

As stated before, despite that the channel estimation and feedback modules are closely related, all the previous CSI feedback schemes assume that the downlink channels are accurately known without channel estimation. 
To realize the complete CSI acquisition process and make the CSI feedback network robust to the errors generated by channel estimation, a joint channel estimation and feedback framework should be constructed. 

In this paper, a comprehensive research is carried out to establish a joint channel estimation and feedback framework for the FDD massive MIMO systems based on DL techniques. The main contributions of this paper are summarized as follows.
\begin{itemize}
	\item The DL based joint channel estimation and feedback framework of downlink channels in FDD massive MIMO systems is proposed in this paper.
	The framework is the first to the best of our knowledge. Two networks are constructed to perform explicit and implicit channel estimation and feedback, respectively. The Channel Estimation and Feedback network (CEFnet) employs a lightweight CNN structure to explicitly obtain the refined estimation of channels and utilizes a Denoising AutoEncoder (DAE) structure to compress and reconstruct the noisy channel matrices. The other pilot compression and feedback network (PFnet) compresses and sends back the pilot information directly to the BS without estimating the channels.

	\item A multi-Signal-to-Noise-Ratios (SNRs) training technique is proposed to cope with multiple SNR cases so that the construction of multiple individual models for each single SNR can be avoided, which significantly reduces the storage space and makes the trained network robust to channel noise. Moreover, quantization module is enrolled into the whole network to generate data-bearing bitstreams and observe the robustness of the two networks to the quantization distortion.
	
	\item Performance analysis of the two proposed networks is provided. Both networks demonstrate excellent reconstruction capacity in that the CEFnet works a little better than PFnet but PFnet generates fewer parameters that need storing than CEFnet. Moreover, the two networks are also proved to be robust to the quantization errors and noise.
	
\end{itemize}

The remaining part of our paper is organized as follows. In Section \ref{section: system model}, a massive MIMO system along with the channel estimation and channel feedback is first introduced. In Section \ref{section: network structure}, two joint channel estimation and feedback frameworks and the quantization technique are described in detail, followed by the corresponding training strategy of each network. Section \ref{section: results} presents simulation results of the proposed networks and investigates the robust performance of the proposed schemes. Section \ref{section: conclusion} finally concludes the paper and points out the future work and major challenges of the DL-based CSI acquisition.

\section{System model}
\label{section: system model}
 
In this section, we will first introduce the massive MIMO system and then describe the channel estimation and feedback process, respectively.
\subsection{Massive MIMO system}
Consider the downlink of a massive MIMO system operating in FDD mode. Orthogonal Frequency Division Multiplexing (OFDM) with $\tilde K$ subcarriers is adopted. The BS is equipped with $N_t(\gg1)$ transmit antennas in the form of Uniform Linear Array (ULA). The user is deployed with a single receiver antenna. 
In the downlink phase, the corresponding received signal component transmitted from the $i$-th transmit antenna can be denoted as,
\begin{equation}
	\mathbf{y}_i = \mathbf{X}_i \tilde{\mathbf{h}}_i \in\mathbb{C}^{\tilde K\times1}
\end{equation}
where $\mathbf{X}_i \in\mathbb{C}^{\tilde{K}\times \tilde{K}}$ is the diagonal matrix of the transmitted signal from the $i$-th transmit antenna, while  $\mathbf{\tilde{h}}_i$ $\in\mathbb{C}^{\tilde{K}\times1}$ represents the channel frequency response vector. 

Then the complete received signal from all $N_t$ transmit antennas can be denoted as,
\begin{equation}\label{receivedSignal}
	\mathbf{y} = \sum_{i=1}^{N_t}\mathbf{X}_i \tilde{\mathbf{h}}_i+\mathbf{n}
\end{equation}
where $\mathbf{n}$ $\in\mathbb{C}^{\tilde K\times1}$ represents the Additive White Gaussian Noise (AWGN) vector. Then, the channel matrix $\tilde{\mathbf{H}} \in\mathbb{C}^{\tilde K\times N_t}$ in the frequency-spatial domain can be obtained by stacking $\tilde{\mathbf{h}}_i$ in the spatial domain as $\tilde{\mathbf{H}}=[\tilde{\mathbf{h}}_1, \tilde{\mathbf{h}}_2, \ldots, \tilde{\mathbf{h}}_{\tilde N_t}]$.

\subsection{Channel estimation}
Attaining the estimated downlink channels is an indispensable prerequisite of the downlink channel feedback. 
To track the instantaneous change of the channel, especially the fading channels, pilot symbol-based channel estimation usually exhibits an excellent performance. 

Since all subcarriers are assumed to be orthogonal, the pilot symbols in $P$ equi-spaced pilot subcarriers can be denoted as a diagonal matrix $\mathbf{X}_p\in\mathbb{C}^{P\times P}$. To obtain the LS-estimated channel response vector between the receive antenna and the $i$-th transmit antenna, the corresponding optimization problem can be denoted as,
\begin{equation}
	\tilde{\mathbf{h}}_p^{LS} = \arg \min\limits_{\tilde{\mathbf{h}}_p} \| \mathbf{y}_p-\mathbf{X}_p \tilde{\mathbf{h}}_p \|_2^2,
\end{equation}
where  $\| \cdot \|_2$ is the Euclidean norm, $\mathbf{y}_p \in\mathbb{C}^{P\times P}$ is the received pilot symbol corresponding to the transmitted pilot symbol $\mathbf{X}_p$. By solving the optimization problem, the channel vector can be estimated, that is $\tilde{\mathbf{h}}_p^{LS}=\mathbf{X}_p^{-1}\mathbf{y}_p$.  Then, interpolation will be conducted to obtain the channel responses at other subcarriers. Since the LS algorithm is easy to operate and requires little computation, it is widely used and will be applied in our framework to obtain the initial estimation.

\begin{figure*}[htb]
	\centering     
	\includegraphics[width=0.85\linewidth]{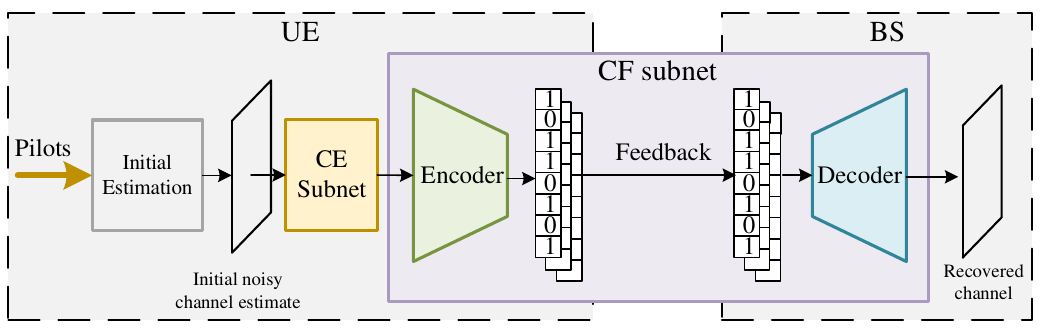}    
	\caption{\label{framework1} Illustration of CEFnet, including CE subnet and CF subnet with an encoder and a decoder. The CE subnet and encoder at the UE obtains CSI and compresses them to codewords which are then quantized into bitstreams and sent back through the uplink.
		The decoder at the BS recovers CSI from feedback bitstreams.}
\end{figure*}

Pilot design is another essential procedure to estimate MIMO channels.
Equi-spaced orthogonal pilots with equi-power have been proved to be the optimal scheme and have been widely used in MIMO systems \cite{minn2006optimal}. Through pilots, the complete channel estimation, $\tilde{\mathbf{H}}_p^{LS} \in \mathbb{C}^{P\times N_t}$, in the frequency-spatial domain can be obtained. However, the increase in the number of transmit antennas will lead to the increase in the pilot overhead, which is tolerable in traditional MIMO systems but is not in massive MIMO because of the huge number of antennas at the BS \cite{rusek2012scaling}. Since there are already many methods, especially these DL ones, that can efficiently address the issue, our research will focus mainly on the channel estimation and feedback scheme.

\subsection{Channel feedback}
In FDD systems, after the UE receives the pilot symbols and estimates the downlink channel matrix, it will send the estimated downlink CSI back to the BS so that the BS can design the corresponding precoding vectors to eliminate interference among users and improve communication quality \cite{9279228}. Since the CSI matrix $\tilde{\mathbf{H}}\in\mathbb{C}^{\tilde K\times N_t}$ contains a total of $2\tilde K N_t$ components, it will cause a heavy feedback overhead in massive MIMO systems. It is not desirable in practical systems, therefore lots of literature has investigated the compressibility of CSI matrix to reduce feedback parameters.

The length of channel impulse response is usually limited, which means the delay spread caused by different arrival times of multiple paths is within a certain period of time. Therefore, the channel vectors show sparsity in the delay domain, only with a few non-zero components.
Moreover, it is proved in \cite{wen2015channel}, as the number of transmit antennas, $N_t$, increases towards infinity, the channel matrix becomes sparse in anglular domain.
However, in practical systems, the number of antennas cannot be infinitely large. Therefore, the channel vectors in the angular domain can only be approximately sparse with few large components and most near-zero components. 
By taking the DFT of the row and column vectors of the CSI matrix $\tilde{\mathbf{H}}$, the approximately-sparse channel matrix $\mathbf{H_{\rm da}}$ in the delay and angular domains can be obtained through a two-dimensional DFT as,
\begin{equation}
	\mathbf{H_{\rm da}} = \mathbf{F}_{\rm d} \tilde{\mathbf{H}}\mathbf{F}_{\rm a}
\end{equation}
where $\mathbf{F}_{\rm d}$ is a $\tilde{K} \times \tilde{K}$  DFT matrix and $\mathbf{F}_{\rm a}$ is a $N_t \times N_t$ DFT matrix. Overall, the sparsification of the channel matrix is a critical prerequisite to reduce the feedback overhead. Due to the limited duration of channel impulse response, the truncation of the channel matrix is first conducted by only reserving the first $K$ rows. Then, the truncated channel matrix $\mathbf{H}$ is vectorized into a sparse vector $\mathbf{h}={\rm{vec}}(\mathbf{H})\in \mathbb{C}^{N\times1} (N=K N_t)$ and compressed by mapping onto a lower-dimensional subspace. The compression process of the channel matrix can be denoted as,
\begin{equation}
	\mathbf{s} = f(\mathbf{h})
\end{equation}
where the function $f(\cdot)$ compresses the $N$-dimensional vectors into $M$-dimensional ones, with a compression ratio of $M/N<1$. The compressed vectors are then transmitted back to the BS. Recovering $\mathbf{h}$ from $\mathbf{s}$ is an under-determined problem but the sparsity property of $\mathbf{h}$ makes it possible to obtain $\mathbf{h}$ based on compressive sensing theory by solving the following constrained optimization problem.
\begin{equation}
	\begin{aligned}
		&\mathbf{\hat h}={\arg\max} \|\mathbf{s}\|_0 \\
		&s.t. \quad \mathbf{s} = f(\mathbf{h})
	\end{aligned}
\end{equation} 

Traditional CS algorithms rely heavily on the prior sparsity assumption of the channel structure, while CSI matrix is only approximately sparse in the delay angular domain.
Therefore, current literature has been making use of the powerful optimization and fitting ability of DL technology to fully learn the knowledge of channel structure through neural network and obtain better reconstruction performance.

\section{DL-based joint channel estimation and feedback framework}
\label{section: network structure}

In this section, two joint channel estimation and feedback networks will be developed, where quantization method will also be discussed. Training strategy of each network will be described at the end of this section.

\begin{figure}[t]
	\centering     
	\includegraphics[width=0.85\columnwidth]{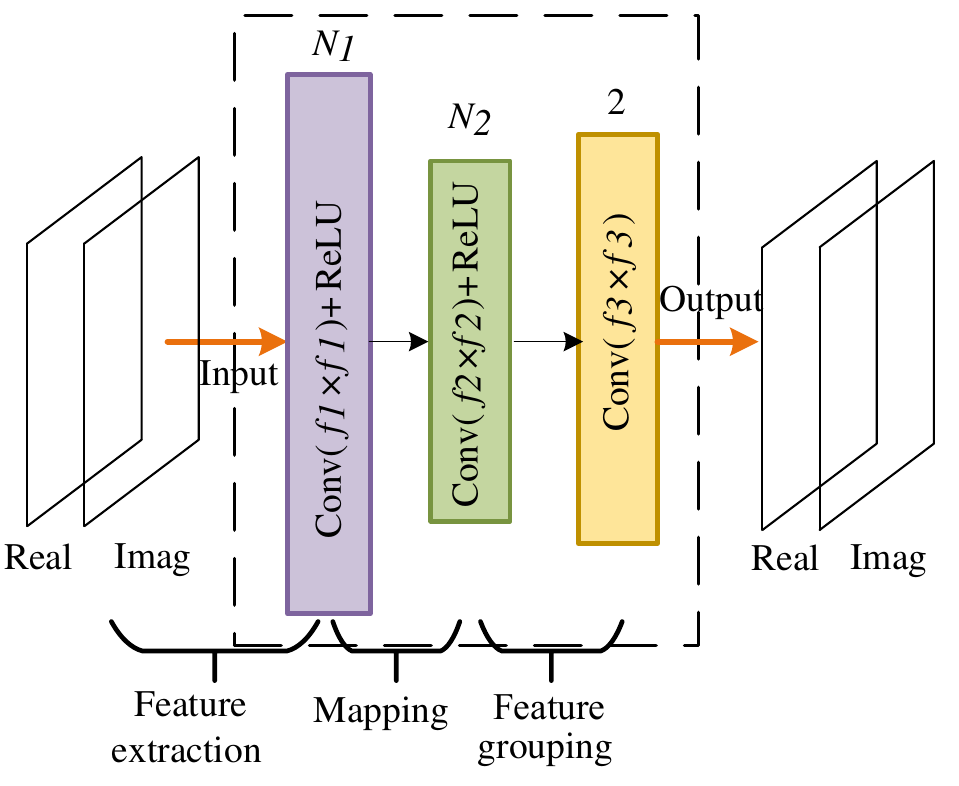}    
	\caption{\label{CEnet}The three-layer CE subnet, with each layer adopting a different-sized filter to perform the specific operation as illustrated.} 
\end{figure}

\begin{figure*}[t]
	\centering     
	\includegraphics[width=0.85\linewidth]{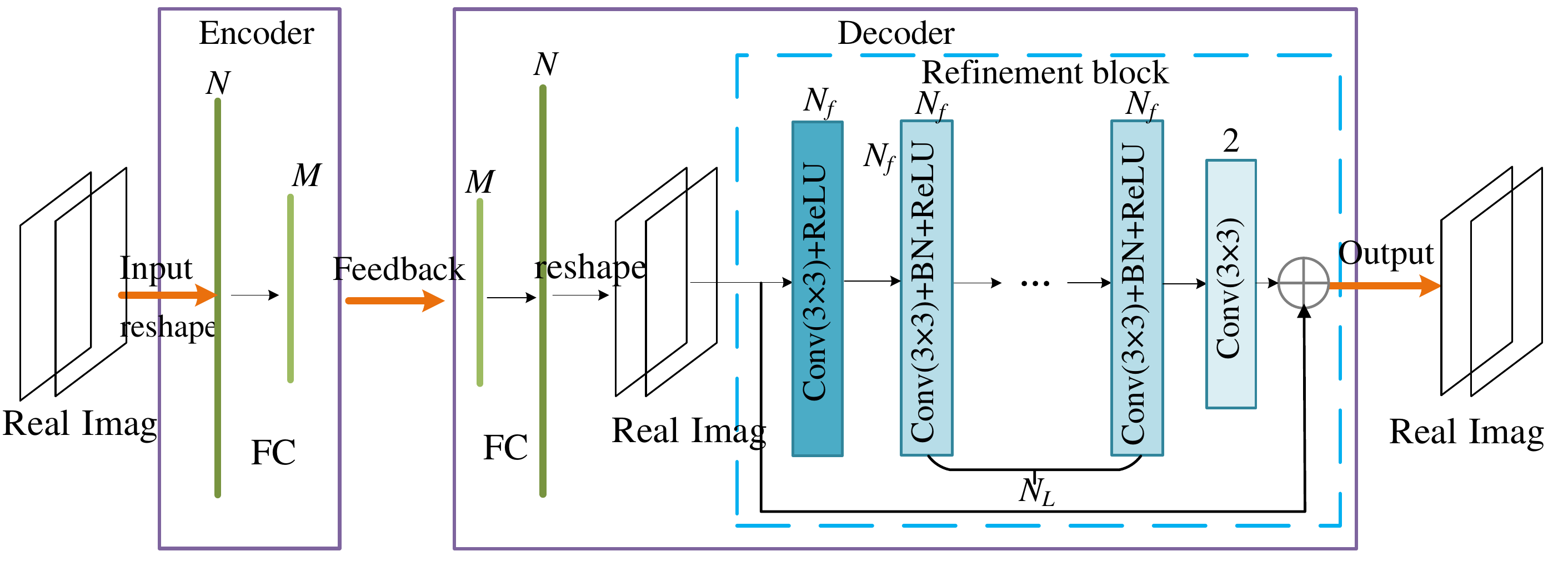}    
	\caption{\label{CFnet}The CF subnet, with the FC layer at the UE compressing the estimated CSI and that at the BS decompressing it. The refinement block based on deep residual network refines it to obtain the final recovered CSI.} 
\end{figure*}

\subsection{Channel estimation and feedback network (CEFnet)}

The whole architecture of the proposed channel estimation and feedback network, called CEFnet, is illustrated in Fig. \ref{framework1}.
In CEFnet, the channel estimation and feedback are successively realized by NNs, which are introduced in Sections \ref{CEsub} and \ref{CFsub}, respectively.
\subsubsection{Channel Estimation (CE) subnet}
\label{CEsub}

Pilot symbol based channel estimation method is adopted to track the instantaneous channel variation.
Due to the special structure of the comb pilots, the receiver only extracts pilot symbols at the pilot locations in frequency domain and then calculates the channel frequency response at the pilot locations. 
Operations (such as interpolation or transformation) are used to obtain the estimated channel frequency response at the remaining locations.
Conventional pilot-based channel estimation methods can be considered as down-sampling, noise-addition, and interpolation of the original channel matrix according to the analysis above. 
In comparison, a Low-Resolution (LR) image is also considered to be derived from a perfect High-Resolution (HR) image after a series of down-sampling and noise interference, based on the imaging principle. 
Image super resolution is to recover the HR images from the LR images, which is similar to the channel estimation optimization process in the theoretical sense.

In recent years, DL has been widely used to establish the nonlinear mapping from an LR image to an HR one to perform Single Image Super Resolution (SISR) \cite{yang2010image,7780551}. 
To apply the DL based SISR methods for channel estimation in our study, the channel matrix is regarded as a two-dimensional image. 
The receiver first estimates the channel by the conventional LS algorithm as an initial channel estimation. 
The estimated channel can be considered as an LR image, with the corresponding ideal channels as the HR one. 
The channel matrix in the space-frequency domain will first be transformed to the angle-delay domain for the compatibility with the following feedback network. 
Due to the limited storage capacity and computation complexity of UEs \cite{wang2021survey}, it is more practical to deploy a light-weight neural network at the UEs. 
Therefore, we adopt a three-layer Super Resolution Convolutional Neural Network (SRCNN) to estimate the channel in our study, which was proposed in \cite{yang2010image} to handle the image super resolution problem.

 The existing DL toolboxes, such as TensorFlow and PyTorch, are designed for real matrix processing, and the complex matrix cannot be directly processed using these toolboxes.
Therefore, following the most existing DL-based CSI feedback works \cite{8322184}, we split the complex channel into two parts (i.e., real and imaginary parts) and stack these two parts on the third dimension \cite{9419066}.
Therefore, the input and output of the networks are two real ``images'' in this work. 
	
As shown in Fig. \ref{CEnet}, the CE subnet is composed of three layers, with each layer performing one specific operation. 
The first layer employing $N_1$ filters of size $f_1\times f_1$ is to extract the raw features of the estimated channel. 
The second layer with $N_2$ filters of size $f_2\times f_2$ $(f_2=1)$, similar to a fully-connected layer, builds the nonlinear mapping relationship between the raw features and the features extracted from the ideal channel. 
Lastly, the mapped features are weighted and grouped to reconstruct the ideal channel matrix by two convolutional filters of size $f_3\times f_3$. 

The Rectified Linear Unit (ReLU), $\rm {ReLU}=\max(x, 0)$, is applied to the output of the first two layers. 
The output of the original network in \cite{yang2010image} has a smaller size than the input. 
To ensure that our output has the same size as the input, we adopt zero padding in every layer during training.

To measure the information retrieval capacity of a convolutional network, the receptive field metric is used to represent the extent to which each pixel in the output image can perceive the original image. 
It can also be taken as the area of each pixel in the output feature map mapped to the original image \cite{luo2017understanding}. 
The reason why the output cannot always perceive all the information of the original image is that the convolutional layer and pooling layer are commonly used and layers are locally connected. 
The larger the receptive field is, the more the information of the original image the network can obtain, which also means that it may contain more global and higher semantic features. 
Therefore, the size of the receptive field can be used to roughly judge the context retrieval capacity of each layer. 
The receptive field of the $l$-th layer can be calculated as follows,
\begin{equation}
	R_l = (R_{l+1}-1)\times s_l+f_l
\end{equation}
where $R_{l+1}$ is the receptive field of the $(l+1)$-th layer, $s_l$ and $f_l$ are the stride and filter size of the $l$-th layer respectively.
According to the above equation and the stride of each layer is 1, the receptive field of the CE subnet can be calculated as.
\begin{equation}
	\label{cerf}
	R_{\rm CEnet} = f_1+f_2+f_3-2
\end{equation}
The calculation result of the receptive field in (\ref{cerf}) will help guide the design of our convolutional network module in our simulation to ensure a stronger information retrieval capacity.

\subsubsection{CSI Feedback (CF) subnet}
\label{CFsub}
As mentioned before, existing DL-based channel feedback frameworks ignore the channel estimation process and directly adopt the autoencoder architecture, of which the input and output are the same ideal channel matrices. 
Since channel estimation has been considered in our proposed framework, the input to the feedback network cannot be regarded as the ideal channel matrix any more. 
The estimated channel should be delivered to the cascaded channel feedback subnet as the input. 
In this way, it can be assumed that the feedback network takes the estimated channel as the noisy case of the ideal channel, which means the feedback network has now taken on a DAE architecture \cite{vincent2008extracting} instead of autoencoder architecture.

To adapt the structure to accomplish the newly-triggered denoising task and meanwhile improve the reconstruction performance, an advanced and high-performance deep residual network in image restoration field \cite{7780551,zhang2017beyond} called VDSR, is introduced into the decoder of our feedback subnet, which is regarded as the refinement block of the decoder. 
The whole architecture of the CF subnet is illustrated in Fig. \ref{CFnet}.

At the UE side, the complex channel matrices should first be divided into the real part and imaginary part since neural networks can only handle real values generally. 
Then, the channel matrices are flattened into a vector of size $N$. 
The encoder at the UE utilizes an FC layer with a Batch Normalization (BN) layer and the tanh activation function to compress the sparse channel vector in the angle-delay domain into a codeword, which is a vector of size $M (M\ll N)$.

At the BS side, the received codeword is first decompressed by an FC layer and then reshaped to the matrix format to obtain an initial coarse estimate of the channel. 
In the next stage, a deep residual CNN is deployed to refine the initial CSI and it is regarded as the major reconstruction module of the decoder. 
The deep residual network contains $L$ layers, where all $N_L(=L-2)$ intermediate layers are set to have the same type: employing $N_f$ filters of the size $3\times3\times N_f$ followed by a BN layer and the ReLU activation function, where each filter performs the convolutional operations on a $3\times3$ matrix region across $N_f$ feature maps of the previous layer. 
The first layer employs $N_f$ filters of size $3\times3\times2$ followed by ReLU activation function, as it operates on the input image. 
The last layer, used for the final reconstruction, consists of 2 filters of size $3\times3\times N_f$. 
The reason why we employ stacked $3\times3$ filters instead of larger-size filters is that large convolutional filters cannot always lead to better performance. 
For example, experiments have revealed that two stacked layers with $3\times3$ filters use fewer parameters, require less computation, and usually perform better than a convolutional layer with a $5\times5$ filter. 

Since filters of the same size $3\times3$ are used for all layers of the reconstruction subnet, it can be calculated that the receptive field of the last layer is of size $3\times3$ and the size of the receptive field for the preceding layers will be doubled in both height and width with the number of layers increasing. 
Therefore, for a network with $L$ layers, the receptive field of the CSI compression and feedback subnet can be calculated as
\begin{equation}
	R_{\rm{CFnet}} = (2L + 1)\times(2L + 1)
\end{equation}
where the size of the receptive field is proportional to the number of layers and can be set to make sure that all elements of the original inputs are exploited to help reconstruct the channels.

\subsection{Pilot compression and feedback network (PFnet)}
\begin{figure*}[t]
	\centering     
	\includegraphics[width=0.65\linewidth]{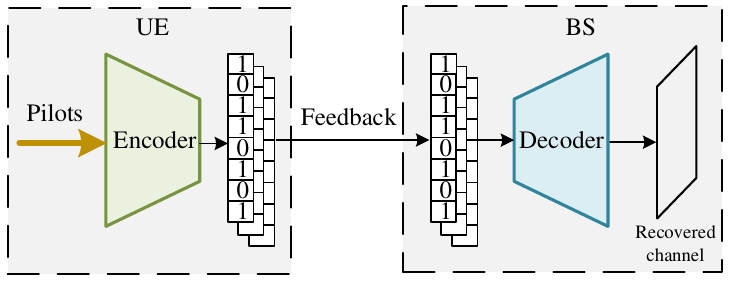}    
	\caption{\label{framework2} Illustration of PFnet, including an encoder and an decoder. The encoder at the UE compresses input pilots to codewords and quantizes them into bitstream and sent back through uplink.
		The decoder at the BS recovers CSI from feedback bitstreams.}
\end{figure*}

Given that the downlink channel estimated at the UE has been compressed for feedback, we can directly feed back the received pilot signals to the BS without channel estimation, which can also reduce the storage space and the computational complexity of the network at the UE.
In this case, the direct pilot compression and feedback network, called PFnet, is proposed, whose input and ideal output are the pilots and ideal channel matrices, respectively. 
It is found that apart from the difference between the inputs of this network and the channel feedback subnet, the two networks perform nearly the same operations: compression, feedback, and reconstruction. 
Therefore, it is convenient and valid to reuse the encoder and decoder of the CF subnet as the architecture of the proposed direct pilot feedback network, as shown in Fig. \ref{framework2}.

To ensure the compatibility of the network with pilots as input, it is indispensable to transform the pilots in the space-frequency domain into the angle-delay domain after the UE extracts the pilot symbols from the received symbols and sets the remaining symbols to zero. 
It is worth mentioning that the pilots can be taken as the weighted channel matrices in this case, which means that pilots also inherit the sparsity in angular-delay domain and make the pilots compressible.
Compared with CEFnet with explicit channel estimation, PFnet can be seen as the implicit channel estimation and feedback.

\subsection{Quantization module}
The compression operation at the UE should be regarded as a reduction of dimension. 
To produce a data-bearing bitstream from compressed codewords, quantization is indispensable in practical communication systems.

To adopt the most suitable quantizer to achieve the minimal performance loss, it is necessary to find out the distribution of the compressed codewords. 
As studied in \cite{8969557}, the values of the majority of components in the codewords are small and close to zero, which makes uniform quantization not suitable to be utilized in this case. 
Therefore, a non-uniform quantizer is utilized to quantize most small values with small spacing and few large values with large spacing to alleviate the quantization distortion caused by uniform quantization. 
A non-uniform quantizer is generated from a uniform one by introducing a nondecreasing and smooth companding function. 
In our study, to investigate the influence of quantization on the whole network, the adaptive-$\mu$ law non-uniform quantizer in \cite{8972904} is adopted. The corresponding companding function can be formulated as
\begin{equation}
	\Phi (x) = \left\vert \frac{{\rm ln}(1+\mu x)}{\rm ln(1+\mu)} \right\vert, 0 \leq x \leq 1
\end{equation}
where $x$ still denotes the normalized value of the codewords, and $\mu$ is the controlling parameter of the companding degree, which is a key parameter to achieve the optimal quantization results. The controlling parameter should be adjusted according to the different distribution characteristics of the input data.
Specifically, a standard method of generating a non-uniform quantizer from a uniform one is to input signals into a companding function and then quantize them by a uniform quantizer.
More details about the $\mu$-law quantization can be found in \cite{8969557}.

The adverse effects of the quantization process on the reconstruction performance of our proposed network will be further discussed in the simulation section.

\subsection{Training strategy}
\subsubsection{CEFnet training  steps}

Given that CEFnet is composed of two subnets (CE subnet and CF subnet), the two subnets should be trained separately. In the first training phase, we train the CEnet with the interpolated noisy channel matrix as input data and the ideal channel matrix as target output. The parameters of the network kernels are denoted as $\Theta_{\rm CE}$. The loss function is set to be the Mean-Squared Error (MSE) between the network output and the target output, which can be expressed as
\begin{equation}
	\label{LossCE}
	L_{\rm CE}(\Theta_{\rm CE})=\frac{1}{T} \sum\limits_{i=1}^{T} \|f_{\rm CE}(\mathbf{H}_{\rm{est},i}; \Theta_{\rm CE}) - \mathbf{H}_i\|_2^2
\end{equation}
where $T$ is the number of training samples, $\| \cdot \|$ is the Euclidean norm, $\mathbf{H}_{\rm{est},i}$ is the initial interpolated estimate of channel matrix and $\mathbf{H}_i$ is the ideal channel matrix.
After training, all interpolated noisy channel matrices will be input to the trained network and the output should be stored locally and then delivered to the cascaded CF subnet as input. 
The training strategy of the CF subnet is similar to CsiNet, whose encoder and decoder is trained jointly in an end-to-end approach. 
The parameters of the second subnet are denoted as $\Theta_{\rm CF}$. 
The loss function is also the MSE, which can be expressed as
\begin{equation}
	\label{LossCF}
	L_{\rm CF}(\Theta_{\rm CF})=\frac{1}{T} \sum\limits_{i=1}^{T} \|f_{\rm CF}(\mathbf{H}_{\rm{CE},i}; \Theta_{\rm CF}) - \mathbf{H}_i\|_2^2
\end{equation}
where $\mathbf{H}_{\rm{CE},i}$ is the output matrix of CEnet.

\subsubsection{Multi-SNRs}
According to the exsiting research, the performance of the channel estimation algorithm is sensitive to the surrounding environments, which means there exists a close connection between the estimation accuracy and the SNR of the received signal. 
In general, the estimation accuracy improves as the SNR increases. 
Therefore, SNR is also an essential variable that will affect the performance of our channel estimation subnet. 
In most existing cases, one network is trained for one single SNR variable and is supposed to work well only with the specified SNR. 

However, in practical communication systems, SNR varies in a range. 
If there is a non-negligible disparity between the SNR of the input data and the SNR under which the model is trained, it is obvious that the network performance will deteriorate. 
Therefore, if a new SNR is under test, a new model has to be trained. 
However, if we separate the SNR range into multiple equi-spaced SNR levels of interest and train the network under these SNR levels to relieve the problem, it turns out that a number of SNR-specific networks are supposed to be trained and stored. 
Assuming that all the networks are considered, the number of the network parameters will be inevitably large, thus taking up a certain amount of storage space of the UE. 
In addition, training multiple individual models for all possible scenarios to adapt to multiple SNR variables is impractical and inefficient. 

For this reason, we need to figure out an effective way to save and restore the networks for use. 
A multi-SNRs scheme is proposed to train the network with multi-SNR input mixed. 
This strategy can be regarded as a novel training method and does not need to change the original networks, including the input, output, and architecture.
Moreover, the training process is the same as the above network except that the training dataset is composed of mixed datasets for all the predefined SNR levels.
Since the estimation accuracy tends to be low when SNR is small, the datasets under small SNRs will have a higher proportion in the final dataset to further boost the performance.

\section{Simulations and results}
\label{section: results}
After introducing data generation and hyperparameter setting, we will present our simulation results in this section.
\subsection{Data generation}
The channels in our simulation are generated by the COST 2100 model \cite{liu2012cost}, which is a widely used Geometry-based Stochastic Channel Model (GSCM) for MIMO systems. 
Two typical scenarios of communication systems are considered: one is the indoor scenario at 5.3GHz, and the other is the outdoor semi-urban scenario at 300MHz. 
In the indoor scenario, the BS is centered at a $20m\times20m$ square region while the UE is moving randomly inside the square. 
In the outdoor scenario, the BS is located at the center of a much larger square area with the side length of $400m$. The BS is deployed with the Uniform Linear Array (ULA) with the number of transmit antennas, $N_t = 32$. The UE has a single receiving antenna. The system is operated in OFDM mode with $\tilde{K}=256$ subcarriers. The channels are sampled every time the UE is moving to a new location. The other channel parameters are set as the default settings in \cite{liu2012cost}. 

The sampled channel matrices are transformed into the angular-delay domain by the 2D-DFT. In the delay domain, the channels are truncated to only reserve the first 32 columns. Therefore, the size of the channel matrices is $32\times32$. Then, the transmitted pilot symbols are generated by placing BPSK-modulated pilots into the equi-spaced subcarriers of each OFDM symbol and are orthogonal to distinguish among different transmit antennas. The received signals are obtained based on equation (\ref{receivedSignal}) with white Gaussian noise added. After transformed into the frequency domain, the received pilot symbols are extracted from the received symbols to generate the needed datasets for simulation use. 
The number of pilot subcarriers in each OFDM symbol depends on the length of the channel impulse response (channel delay) since the pilot length must be larger than that of the channel impulse response so that the channel is possible to be estimated. 

\begin{figure*}[t]
	\subfigure[indoor\_CR4\_P16]{
		\begin{minipage}{0.5\linewidth}
			\centering
			\includegraphics[width=0.95\linewidth]{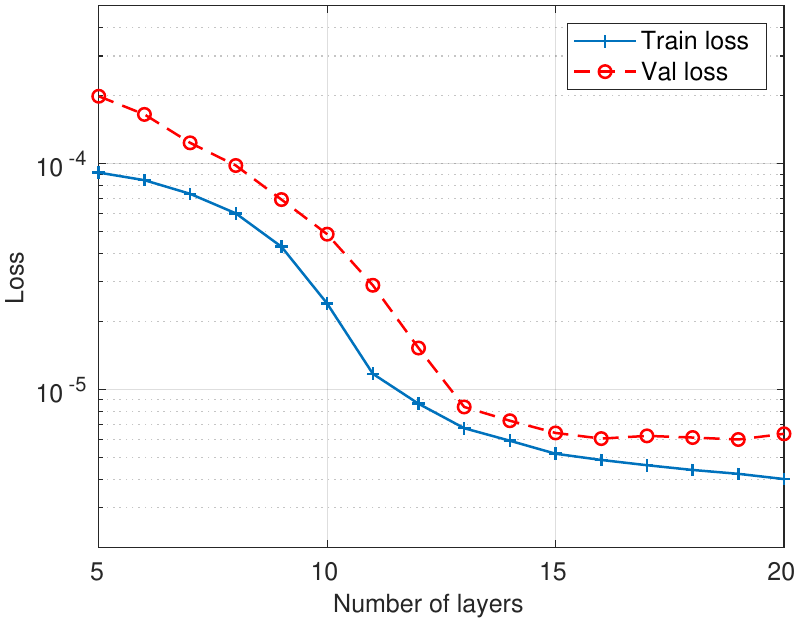}
	\end{minipage}}
	\subfigure[outdoor\_CR4\_P32]{
		\begin{minipage}{0.5\linewidth}
			\centering
			\includegraphics[width=0.95\linewidth]{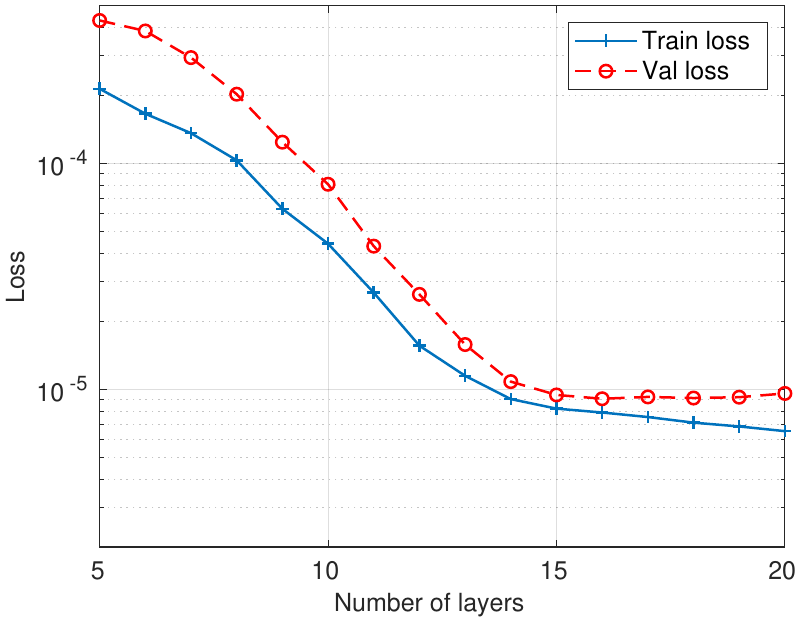}
		\end{minipage}
	}
	\caption{The learning curve of two specific scenarios based on grid search. (a)The indoor scenario with $CR$=4 and $P$=16. (b)The outdoor scenario with $CR$=4 and $P$=32. }\label{lc}
\end{figure*}

The final input samples of CEFnet are the channel matrices estimated by the LS algorithm and interpolated based on pilot symbols while the input of PFnet is the extracted pilot symbols. The target output of the two networks is the ideal channel matrices. Both the input and output need transforming into the angular-delay domain before inputting the network to exploit the sparsity. 

A standard dataset is composed of the input samples and the corresponding target output.
To achieve the optimal training results, the datasets should be shuffled and divided into three parts: training dataset for updating the network parameters to make the model converge, validation dataset for tracking and estimating the generalization performance of the network, and testing dataset for evaluating the performance of the final trained model. Note that the testing dataset should not be included in the training or validation dataset to avoid data snooping. The numbers of samples in the three sets are 100,000, 30,000 and 20,000, respectively.

In addition, the performance metric for the whole simulation is the Normalized MSE (NMSE) between the output of the network and the true channel matrices, which can be calculated as follows
\begin{equation}
	\rm NMSE =   \rm E \{ \frac{\|  \mathbf{H} -   {\mathbf{\hat H}}\|^2_2}{\|\mathbf{H}\|^2_2 }\}
\end{equation}

\subsection{Network setting}
Simulations are operated on an NVIDIA DGX-1 workstation with 8 GPUs using Keras library, with Tensorflow as the backend.
The parameters of the two networks are initialized by the Glorot uniform initializer and updated by the Adam optimizer. The batch size is 200. The initial epoch is set to be 200 while the early-stopping technique is employed to avoid wasting time and resources since it can interrupt training when it detects no decline in the validation loss after a large enough number of epochs and the last best model will be stored according to the training log. The initial learning rate is 0.01 and will be reduced by half if the validation loss does not experience a decline within a number of epochs.

In the feedback process, the channel matrices are compressed and the Compression Ratio ($CR$) exerts a significant influence on the reconstruction performance, which can be defined as follows
\begin{equation}
	CR = \frac{2KN_t}{M}
\end{equation}
where $M$ is the length of the codeword. In our simulation, the $CR =$ 4, 8, and 16, and the corresponding $M =$ 512, 256 and 128, respectively.

In section \ref{section: network structure}, the general architecture of the two networks has been briefly described while the specific network settings will be introduced in this section. First, the channel estimation subnet employs 64 convolutional filters of size $9\times9$ in the first layer, 32 filters of size $1\times1$ in the second layer and 2 filters of size $5\times5$ in the final layer. Therefore, the receptive field of the channel estimation subnet is $13\times13$. Next, the number of output neurons in the one-layer FC network of the encoder is equal to the aforementioned codeword length M. So is the number of input neurons in the one-layer FC network of the decoder. In the last part, the reconstruction subnet of the decoder employs a convolutional layer with 64 filters of size $3\times3$ followed by the ReLU activation function in the first layer and a layer with 2 filters of size $3\times3$ in the last layer. Grid search is adopted to obtain the optimal number of the intermediate layers, the results of which are shown in Fig. \ref{lc}. Two specific cases (indoor and outdoor) are considered and experimented on to plot the learning curves.

According to the learning curve in Fig. \ref{lc}, the training loss follows a downward trend as the number of layers increases.
However, the validation loss decreases in the first phase before the number of layers reaches 15 or 16 but it stops going down and even sees a little increasing trend in the following phase. It indicates that when the number of layers increases above 16, the model is so deep and complicated that it has overfitted the training dataset. Based on the grid search results, the best number of layers is decided to be 16. The receptive field is thus $33\times33$ through calculation. Since the size of input matrices is $32\times32$, the network has fully exploited the contextual input information to predict the unknown components of the channel matrices. Despite that the reconstruction is an ill-posed inverse problem, enough neighbor values have been collected and analyzed by the network so that the reconstruction is resolvable, which further proves that 16 layers are the most suitable option.
\begin{table}[t]
	\centering
	\caption{The detailed network setting of the proposed networks}
	\label{network parameters}
	\resizebox{\columnwidth}{!}{%
		\begin{threeparttable}
		\begin{tabular}{c|c|c|c|c}
			\hline\hline
			& Module & Layer &Parameter$^*$ & Location \\
			\hline
			\multirow{11}{*}{\rotatebox{90}{CEFnet} } & \multirow{3}{*}{CE subnet} & Conv\_1+ReLU & $64\times9\times9\times2$ & UE \\
			&  & Conv\_2+ReLU & $32\times1\times1\times64$ & UE \\
			&  & Conv\_3 & $2\times5\times5\times32$ & UE \\ \cline{2-5}
			& \multirow{8}{*}{\begin{tabular}[c]{@{}c@{}}
					CF subnet/\\(PFnet)\end{tabular}} & Reshape & None & UE \\
			&  & FC\_1+BN+Tanh & $N\times M$ & UE \\
			&  & Quantization & None & UE \\
			&  & FC\_2+BN+ReLU & $M\times N$ & BS \\
			&  & Reshape & None & BS \\
			&  & Conv\_4+ReLU & $64\times3\times3\times2$ & BS \\
			&  & Conv\_5$\sim$18+BN+ReLU & $64\times3\times3\times64$ & BS \\
			&  & Conv\_19 & $2\times3\times3\times64$ & BS\\ \hline \hline
		\end{tabular}
	\begin{tablenotes}
		\footnotesize
		\item[*] $N_{\rm out}\times f\times f \times N_{\rm in}$ represents that the input and output feature numbers are $N_{\rm in}$ and $N_{\rm out}$, respectively, and the convolutional kernel size is $f \times f$.
		$N_{\rm in} \times N_{\rm out}$ represents the input and output neuron numbers of the FC layer are $N_{\rm in}$ and $ N_{\rm out}$, respectively.
		
	\end{tablenotes}
\end{threeparttable}
	}
\end{table}
Based on the above discussion, Table \ref{network parameters} demonstrates all the parameter settings and activation functions of the two proposed networks in detail, including the number of layers, filters, and neurons.

\subsubsection{Simulation for the channel estimation subnet only}

For conventional algorithms, the channel can be estimated only when the pilot length is over that of channel impulse response. After the observation of the two scenarios discussed before, the lengths of channel impulse response in the indoor scenario and outdoor scenario are not longer than 8 and 32, respectively. Therefore, in this study, the pilot length for the indoor scenario is set to be 8, 16, and 32, while the pilot length for the outdoor scenario is 32 and 64.

\begin{figure*}[htbp]
	\subfigure[indoor]{
		\begin{minipage}{0.5\linewidth}
			\centering
			\includegraphics[width=0.95\linewidth]{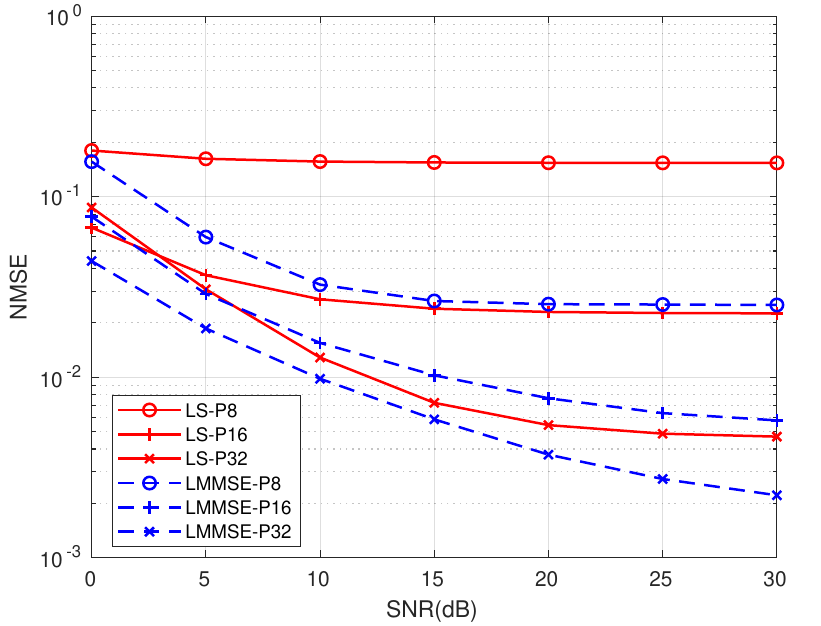}
	\end{minipage}}
	\subfigure[outdoor]{
		\begin{minipage}{0.5\linewidth}
			\centering
			\includegraphics[width=0.95\linewidth]{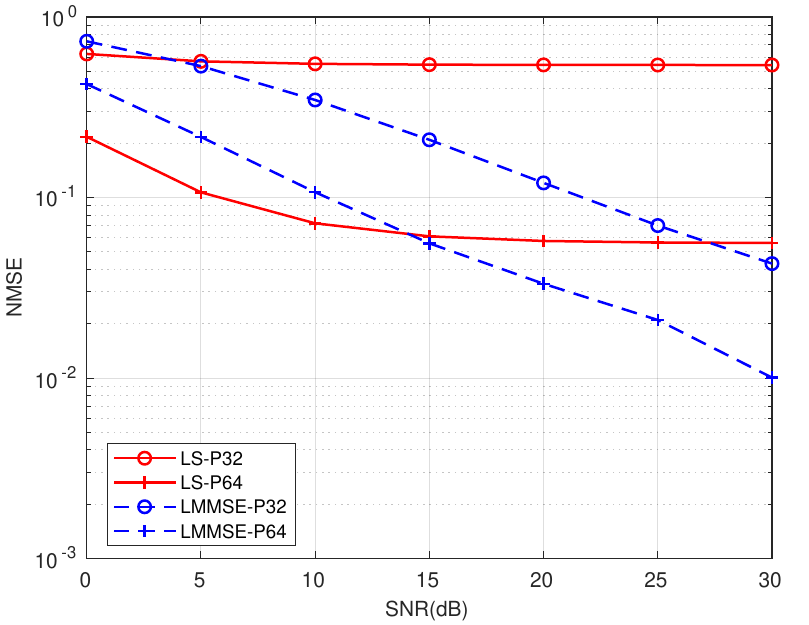}
		\end{minipage}
	}
	\caption{The NMSE ($dB$) performance of the initial channel estimates in different cases based on conventional algorithms.}\label{CE1}
\end{figure*}
The performance of the conventional channel estimation algorithms, such as the LS algorithm and the Linear Minimum MSE (LMMSE) algorithm, is first simulated to serve as a benchmark of the following channel estimation subnet. Fig. \ref{CE1} illustrates the NMSE performance of the interpolated LS algorithm and the LMMSE algorithm under different pilot lengths and different SNRs for the indoor and outdoor scenarios, respectively.

\begin{table}[t]
	\centering
	\caption{The NMSE ($dB$) performance of CE subnet}
	\label{CEcompare}
	\resizebox{0.9\columnwidth}{!}{%
		\begin{tabular}{c|ccc|cc}
			\hline \hline
			&
			\multicolumn{3}{c}{Indoor} &
			\multicolumn{2}{c}{Outdoor} \\ \hline
			&
			P8 & P16 & P32 & P32 & P64 \\ \hline
			LMMSE & -14.87 & -18.10 & -20.08 & -4.59 & -9.70 \\
			LS+interpolation &
			-8.05 & -15.68 & -18.92 & -2.59 & -11.42 \\
			CE subnet &
			-11.26 & -17.09 & -19.51 & -8.39 & -16.43\\ \hline \hline
		\end{tabular}%
	}
\end{table}
From the figure, the overall estimation error is large when conventional algorithms are used. With the increase in SNR or pilot length, the estimation error will decrease. In addition, as the pilot length increases, the channel estimation will be more accurate and the influence of SNR on the estimation error will be more distinct. Then, the estimated channel matrices by the conventional algorithm with SNR = 10 dB in each scenario are collected as the dataset and input into the channel estimation subnet for training. The performance of the subnet under different pilot lengths in each scenario is compared with that of the benchmarks in Table \ref{CEcompare}.

From Table \ref{CEcompare}, in the indoor scenario, the NMSE performance of the neural network-based channel estimation is about 0.58 dB, 1.41 dB, 3.20 dB, lower than the conventional algorithm, when the pilot length is 8, 16, 32, respectively. Therefore, the neural network exerts a significant influence on channel estimation. Furthermore, the shorter the pilot length is,
the stronger the ability of the neural networks becomes to improve the channel estimation accuracy.
Similar trend can be found in the outdoor scenario. However, the overall improvement in the outdoor scenario is more distinct than that in the indoor one. Based on the above analysis, when the condition or environment becomes worse and harsher, the channel estimation subnet tends to get more powerful to obtain a more accurate channel estimation. 

\begin{table}[t]
	\caption{The NMSE ($dB$) performance comparison between the single-SNR model and the multi-SNRs model in both scenarios.}
	\label{multiSNR}
	\centering
	\newcommand{\tabincell}[2]{\begin{tabular}{@{}#1@{}}#2\end{tabular}}
	\resizebox{\columnwidth}{!}
	{
		\begin{tabular}{c|ccccccc}\hline \hline
			\multicolumn{8}{c}{Indoor}\\ \hline
			SNR(dB) & 0 & 5& 10& 15& 20& 25& 30 \\ \hline
			Initial & -13.98 & -16.42 & -17.52 & -17.94 & -18.02 & -18.22 & -18.23 \\
			\tabincell{c}{single-SNR(10 dB)}& 
			-14.55 & -17.42 & -18.87 & -19.50 & -19.61 & -19.86 & -19.87 \\
			multi-SNRs &
			\textbf{-15.91} & \textbf{-18.45} & \textbf{-19.78} & \textbf{-20.52} &\textbf{-20.71} & \textbf{-20.99} &\textbf{-21.00} \\	 [2pt] \hline
			
			\multicolumn{8}{c}{Outdoor}\\ \hline
			SNR & 0&5&10&15&20&25&30\\ \hline
			Initial & -9.32 & -13.26 & -16.02 & -17.39 & -17.97 & -17.97 & -18.39 \\
			\tabincell{c}{single-SNR(10 dB)}& -9.70 & -13.67 & -16.43 &-17.78 & 18.39 & -18.40 & -18.78 \\
			multi-SNRs &
			\textbf{-9.93} & \textbf{-14.01} &\textbf{-16.78} & \textbf{-18.04} & \textbf{-18.61} &\textbf{-18.75} &\textbf{-18.97} \\ \hline \hline
	\end{tabular}}
\end{table}

\begin{table}[t]
	\tiny
	\centering
	\caption{The NMSE ($dB$) performance comparison with previous methods}
	\label{CFnetCompare}
	\resizebox{\columnwidth}{!}{%
		\begin{tabular}{c|c|ccc}
			\hline \hline
			Scenario& {$CR$}   & 4  &  8      &  16    \\ \hline
			&LASSO  & -7.59  &  -4.72  & -2.72 \\
			&  CsiNet &  -17.43 & -13.65 &  -8.79 \\ 
			\multirow{-3}{*}{ Indoor} &
			CF subnet&\textbf{-20.84} &\textbf{-17.90} &\textbf{-13.40}\\ \hline
			&LASSO  &  -5.08  &  -2.97 & -1.01 \\
			&CsiNet &  -8.62  &  -6.38  & -4.12 \\
			\multirow{-3}{*}{ Outdoor} &
			CF subnet &
			\textbf{-9.78} &
			\textbf{-7.08} &
			\textbf{-4.71} \\ \hline \hline
		\end{tabular}%
	}
\end{table}

However, as mentioned before, one model is created for one single case in neural networks. If the cases under all different SNRs and pilot lengths are considered, we need to carry out a total of $7\times3+7\times2=35$ training, and 35 models will be obtained. It is obviously impractical for storage and future use. Therefore, the multi-SNRs technique will be utilized here. In most cases, the received SNR at the UE usually ranges from 0 to 30 dB, so we use 7 SNR levels within the range, 0, 5, 10, 15, 20, 25 and 30 dB. The input samples are changed to be combined by the samples at the 7 different SNR levels. 
Since the smaller the SNR, the worse the channel estimation, the estimation performance of the whole network can be improved by increasing the proportion of the dataset with small SNRs.

To further enhance the influence of mixture, the dataset of small SNRs that occur more often in practical and may lead to worse estimation accuracy are given a larger proportion.
The low SNR scenario occurs often in practical systems and may lead to worse estimation.
Therefore, the dataset with low SNRs should be given a larger proportion to ensure the gain of the proposed mixture strategy.
Therefore, the proportion of 2:2:2:1:1:1:1 is allocated to the SNRs of 0, 5, 10, 15, 20, 25 and 30 dB to generate a new input dataset. Table \ref{multiSNR} compares the NMSE of the multi-SNRs model to estimate the channels under all 7 SNR levels with that of the single-SNR model, which is trained under SNR of 10 dB. The pilot length is 16 and 32 in the indoor and outdoor scenarios, respectively.

\begin{table}[t]
	\centering
	\caption{The NMSE ($dB$) performance of the two proposed networks.}
	\label{finalCompare}
	\resizebox{\columnwidth}{!}{
		\begin{tabular}{c|c|ccc|cc} \hline \hline
			&
			&
			\multicolumn{3}{c|}{Indoor} &
			\multicolumn{2}{c}{Outdoor} \\ \hline
			\diagbox{$CR$}{$P$}&Method & 8 & 16 &32 & 32 & 64 \\ \hline
			&ideal &
			\multicolumn{3}{c|}{{-20.84}} &
			\multicolumn{2}{c}{{-9.78}} \\ \cline{2-7}
			&PFnet &-14.61 & -18.07 &-18.93 & -7.55 & -8.67 \\
			\multirow{-3}{*}{4} &
			CEFnet &
			-16.21 & -19.01 & {-20.31} & -7.72 & {-9.48} \\ \hline
			&ideal &
			\multicolumn{3}{c|}{ {-17.90}} &
			\multicolumn{2}{c}{ {-7.08}}\\ \cline{2-7}
			&PFnet & -11.30 &-13.71 & -15.06 &-4.43 & -5.36 \\
			\multirow{-3}{*}{ 8} &
			CEFnet &
			-13.77 &-16.53 & {-17.65} & -4.45 & {-6.78} \\ \hline
			&ideal &
			\multicolumn{3}{c|}{ {-13.40}} &
			\multicolumn{2}{c}{ {-4.71}} \\ \cline{2-7}
			&PFnet & -8.12 & -9.63 & -10.97 & -3.11 &-3.46 \\
			\multirow{-3}{*}{16} &
			CEFnet &
			-10.69 & -12.53 & {-13.35} & -4.23 & {-4.63} \\ \hline \hline
		\end{tabular}
	}
\end{table}

\begin{table*}[t]
	\centering
	\caption{The NMSE ($dB$) performance of the proposed quantization methods for different scenarios.}
		\begin{tabular}{c|ccc|ccc|ccc}  \hline \hline
			&\multicolumn{9}{c}{Indoor\_P32}  \\  \hline 
			\multicolumn{1}{c|}{$CR$}                                  & \multicolumn{3}{c|}{4} &\multicolumn{3}{c|}{8} & \multicolumn{3}{c}{16}                              \\
			\hline
			\multicolumn{1}{c|}{$B$}                                     & 3  & 4  & \multicolumn{1}{c|}{5}  & 3   &4  &\multicolumn{1}{c|}{5}  &3 & 4 &5\\
			\hline
			\multicolumn{1}{c|}{PFnet}&
			\multicolumn{3}{c|}{-18.94}&
			\multicolumn{3}{c|}{-15.06} &
			\multicolumn{3}{c}{-10.97} \\ \hline
			\multicolumn{1}{c|}{PFnet+Q} &
			-13.32 &-14.93  & {-17.04} &  -10.01& -12.79 & {-14.40}   & -8.79 & -9.37 & {-10.62} \\
			\hline
			\multicolumn{1}{c|}{CEFnet}&
			\multicolumn{3}{c|}{-20.31}&
			\multicolumn{3}{c|}{-17.65} &
			\multicolumn{3}{c}{-13.36} \\ \hline
			\multicolumn{1}{c|}{CEFnet+Q} &
			-14.82 &-16.23  & {-18.65} &  -12.28& -15.90 & {-17.00}   & -11.61 & -12.28 & {-13.03} \\
			\hline \hline
			
			& \multicolumn{9}{c}{Outdoor\_P64}                                                                                                                                                                                                                                                            \\  \hline 
			\multicolumn{1}{c|}{$CR$  }                                                           & \multicolumn{3}{c|}{4}                               & \multicolumn{3}{c|}{8}                               & \multicolumn{3}{c}{16}                               \\
			\hline
			\multicolumn{1}{c|}{$B$}                                    & 3  & 4 & \multicolumn{1}{c|}{5}             & 3       & 4    & \multicolumn{1}{c|}{5}         & 3    & 4 & 5    \\
			\hline
			\multicolumn{1}{c|}{PFnet}&
			\multicolumn{3}{c|}{-8.67}&
			\multicolumn{3}{c|}{-5.37} &
			\multicolumn{3}{c}{-3.46} \\ \hline       
			PFnet+Q & -6.21          & -7.47          & {-8.00}           & -3.38           & -4.32           & {-4.97}        & -2.17           & -2.79          & 	{-3.02}        \\
			\hline                                              
			\multicolumn{1}{c|}{CEFnet}&
			\multicolumn{3}{c|}{-9.48}&
			\multicolumn{3}{c|}{-6.78} &
			\multicolumn{3}{c}{-4.63} \\ \hline                                                         CEFnet+Q & -7.22          & -8.22          & {-9.08}           & -5.22           & -6.00           & {-6.48}        & -3.69           & -4.14          & 	{-4.39}        \\
			\hline \hline
	\end{tabular}
	\label{quanNMSE}
\end{table*}

From Table \ref{multiSNR}, in the indoor scenario and for the same testing data, the performance of the channel estimation subnet trained under a single specific SNR significantly decreases when the SNR of the testing input is smaller than the model’s own SNR. Although the relative performance of the channel estimation subnet trained under multi-SNRs also experiences a slight decrease when the SNR of the testing input is small, the absolute performance is better and the overall network performance is improved compared with the that of the single-SNR model. In contrast with the indoor scenario, the relation between SNR and NMSE becomes a little obscure after going through the neural network in the outdoor scenario. The NMSE does not show an absolute upward trend any more with the SNR increasing. However, under the same testing data, the overall network performance of channel estimation subnet trained under multi-SNRs is improved distinctly although the improvement is a bit unstable. Therefore, the subnet trained under multi-SNRs gains better robustness and can adapt to the task of channel estimation under different SNRs.

\subsubsection{Simulation for the CSI feedback subnet only}

Before the simulation is conducted on the whole network, it is necessary to measure the compression and reconstruction ability of the CSI compression and feedback subnet. Based on our study, CsiNet cannot converge with the noisy input. Therefore, the dataset of the ideal channel matrices as both input and output is utilized here to accomplish the measurement, which is realized by comparing the NMSE performance of our proposed subnet with that of the CsiNet and the state-of-the-art CS-based method, namely LASSO algorithm. Table \ref{CFnetCompare} shows the comparison results under different CRs for the indoor and outdoor scenarios. As shown in the table, the proposed subnet outperforms the CsiNet and the LASSO algorithm in the ability of feature extraction and reconstruction.

\subsubsection{Simulation for the whole network}

Table \ref{finalCompare} compares the performance of the two networks (CEFnet and PFnet).
The ideal case in the table refers to the method that feeds back the perfect downlink CSI with the feedback network proposed in this work.
From the table, the reconstruction accuracy gets higher with the increase in pilot length for both networks. However, the improvement of the performance becomes less distinct when the pilot length gets long enough. For example, in the indoor scenario with CR=4, the NMSE decreases about 3.4 dB when the pilot length increases from 8 to 16, but the decrease is only about 0.9 dB when the pilot length of 16 turns 32. Moreover, the results also show that as the CR becomes larger, it gets much more difficult to reconstruct the channel. 
Be it in the indoor or outdoor scenario, the same trend is found in both CEFnet and PFnet. This may be attributed to the reason that highly accurate reconstruction relies on the information contained in codewords. When the CR is larger, the codewords become shorter and thus less useful channel information can be exploited for recovery. 

To compare the indoor and outdoor scenarios from the two tables, the NMSEs in the outdoor scenarios are generally larger than those indoors. It infers that it is much harder to reconstruct the outdoor channels than the indoor channels, because the outdoor environment is more complex and harsher than the indoor one. 

Then, to compare the CEFnet with PFnet, the ideal case where the input and target output are both ideal channel matrices is also listed in the table to serve as a benchmark. In the indoor scenario, the reconstruction performance of CEFnet is over 1 dB better than that of PFnet under all pilot lengths and all CRs. While in the outdoor, CEFnet outperforms PFnet only by a small margin in most cases, especially when the pilot length is short. Besides, although the NMSE performance demonstrates a predictable gap between the ideal case and the two frameworks, it turns out that CEFnet exhibits a similar performance as the ideal case when the pilot length is long enough, which also demonstrates that CEFnet has a powerful reconstruction ability.

In terms of the complexity of the two networks, CEFnet possesses 14,016 more parameters than PFnet as it includes one more functional subnet. Therefore, the running time of CEFnet excels PFnet by around 20 $\mu$s. 
Given the limitation of the computation of the UE, the networks at the UE are rather simple and composed of three fully connected layers with a few neurons.
Specifically, the running times of the networks at the UE and BS are less than 0.5 ms and 2.5 ms, which can meet the requirement of practical systems.
	
In conclusion, CEFnet and PFnet can well recover the channel from the interpolated noisy pilots although there is a difference in reconstruction performance and complexity between them.

\subsubsection{Simulation for quantization}
Quantization is dispensable to generate the bitstream that can be stored and transmitted. A specific non-uniform quantizer described above is adopted to realize the goal. In this part, the effects of the proposed quantization on the reconstruction will be measured and discussed. CEFnet is taken as the testing object. One case is selected from the indoor and outdoor scenarios, with the indoor case under pilot length of 32 and the outdoor case with the pilot length of 64.

The results in Table \ref{quanNMSE} indicate that the change of quantization bits exerts more influence on the indoor cases than the outdoor ones, because the reconstruction performance shows a significant improvement when quantization bits increase. Moreover, low $CR$ cases are more sensitive to the quantization bits than high $CR$ cases, because the information in codewords is significant for reconstruction in the low CR cases and few errors introduced into the codewords will worsen the reconstruction accuracy. 
Besides, with the quantization bits increasing, the reconstruction performance gets closer to that without quantization as we can imagine, which indicates that quantization does not have much adverse effect on the proposed network as it can be handled at the decoder.

\section{Conclusion}
\label{section: conclusion}
In this paper, we have proposed two DL-based joint channel estimation and feedback networks for the estimation, compression, and reconstruction of the downlink channels in the FDD massive MIMO systems.
Two networks are constructed to achieve the explicit and implicit channel estimation and feedback respectively. The CEFnet adopts a multi-SNRs technique to train the CE subnet to adapt to different SNRs by mixing the input data of different SNRs based on a certain proportion and deploys a deep residual network to reconstruct the channels. The PFnet realizes implicit channel estimation and feedback by directly compressing pilots and sending them back, which also reduces network parameters. The two networks are both trained in an end-to-end manner. Quantization module is also enrolled according to the distribution of codewords to generate data-bearing bitstreams. Simulation results show that the two proposed networks both outperform CsiNet, while the CEFnet exhibits better performance in reconstruction than PFnet with more network parameters. Furthermore, the two networks are robust to different environments and quantization errors.

Although the proposed DL-based joint channel estimation and feedback has shown great potentials in simulation, some extensive challenges are worth further exploring in the future. 
First, the collection of the training channel datasets in practical systems is difficult. DL-based algorithms rely heavily on the training set.
If the quality of the training channels is poor, the performance drop is unavoidable.
Then, the DL-based CSI acquisition requires high computation compared with conventional algorithms.
Therefore, network compression, such as pruning and quantization, is essential.
Moreover, although the proposed methods are robust to different SNRs, the generalization to different environments is poor.
Online learning is essential if the channel environment changes.
Finally, all experiments in this work are based on simulated channels, and the proposed DL-based CSI acquisition should be evaluated using a real-world channel dataset.

%


\bibliographystyle{elsarticle-num}
\bibliography{reference}
\end{document}